\def \cm{~\rm{cm}}
\def \s{~\rm{s}}

\def \K{~\rm{K}}

\def \erg{~\rm{erg}}

\def \kpc{~\rm{kpc}}
\documentclass[12pt,preprint]{aastex}
\usepackage{natbib}

\shorttitle{Pairs of bubbles}
\shortauthors{Soker}
\slugcomment{Draft version of \today}

\begin{document}

\title{BUBBLES IN PLANETARY NEBULAE AND CLUSTERS OF GALAXIES:
INSTABILITIES AT BUBBLE' FRONTS}

\author{Noam Soker\altaffilmark{1}}

\altaffiltext{1}{Department of Physics, Technion$-$Israel
Institute of Technology, Haifa 32000, Israel,
and Department of Phsyics Oranim;
soker@physics.technion.ac.il.}

\begin{abstract}

I study the stability of off-center low-density
more or less spherical (fat) bubbles in clusters
of galaxies and in planetary nebulae (PNs) to Rayleigh-Taylor (RT)
instability. As the bubble expands and decelerates, the interface
between the low-density bubble's interior and the dense shell formed
from the accreted ambient medium is RT-stable.
If, however, in a specific direction the density decreases
such that this segment is accelerated by the pressure inside the
bubble, then this accelerated region is RT-unstable.
The outermost region, relative to the center of the system,
is the most likely to become unstable because there the
density gradient is the steepest.
Using simple analytical analysis, I find that off-center fat
bubbles in PNs are much less stable than in clusters.
In PNs bubbles become unstable when they are very small relative to
their distance from the center; they can be stabilized somewhat
if the mass loss rate from the stellar progenitor decreases for a time,
such that the negative density gradient is much shallower.
In clusters fat bubbles become unstable when their size is comparable
to their distance from the center.
I discuss some implications of this instability in clusters and in PNs.

\end{abstract}


\section{INTRODUCTION} \label{sec:intro}

This is a third in a series of papers that aim at exploring
the nontrivial similarities between pairs of bubbles in
clusters of galaxies and in planetary nebulae (PNs).
The similarities in morphologies were discussed in Soker
(2003a; hereafter paper-I; see also section 5 in Soker 2003c),
where a table comparing X-ray images of clusters with optical images
of PNs is given.\footnote{The images of the objects listed in table 1
of paper-I are summarized in a {\it PowerPoint} file I presented at
the Asymmetrical Planetary Nebulae III meeting (2003), at
\newline
http://www.astro.washington.edu/balick/APN/APN\_talks\_posters.html
\newline
(go the `ppt' file in the ``discussion'' of session 13).}
In clusters, poor galaxy groups, and elliptical galaxies the
bubbles are X-ray deficient bubbles (e.g., Hydra A,
McNamara et al.\ 2000; Perseus, Fabian et al.\ 2000, 2003;
A 2597, McNamara et al.\ 2001; RBS797, Schindler et al.\ 2001;
Abell~4059, Heinz et al.\ 2002; Abell~2052, Blanton et al. 2003;
HCG 62, Vrtilek et al.\ 2002; M84, Finoguenov \& Jones 2001;
NGC 4125 and NGC 4552, White \& Davis 2003).
In PNs these are optical deficient bubbles
(e.g., the Owl nebula [NGC 3587; PN G148.4+57.0],
Guerrero et al.\ 2003;  Cn 3-1 [VV 171; PN G038.2+12.0],
Sahai 2000; Hu 2-1 [PN G051.4+09.6], Miranda et al.\ 2001).
With $\sim 5$ orders of magnitude difference in size,
$\sim 3$ orders of magnitude difference in temperature,
and $>10$ orders of magnitude difference in energy, the
similarities in bubbles' morphologies between some PNs and some
cooling flow clusters of galaxies and poor groups is striking.
In paper-I I also point to similar values of some non-dimensional
quantities between clusters and PNs.
These similarities led me to postulate a similar formation mechanism,
hence strengthening models for PN shaping by jets (not all PNs
are shaped by jets).
I further argue there that the jets were blown by binary companions.

In a second paper (Soker 2003b; paper-II), I derive constraints on
the jet properties for inflating pairs of fat-bubbles, i.e.,
more or less spherical.
In paper-II it was found that for inflating fat bubbles the
opening angle of the jets should be large, i.e., the half opening
angle measured from the symmetry axis of the jets should typically be
$\alpha \gtrsim 40 ^\circ$, or the jets should precess.
(For such wide-opening angle jets, a collimated fast wind [CFW]
is a more appropriate term.)
Narrow jets will form elongated lobes rather than fat bubbles.

In the present paper I study the development of the
Rayleigh-Taylor (RT) instability in the interface between the
hot tenuous bubble interior and the outer dense shell.
My goals are as follows.
\begin{enumerate}
\item Find the conditions for the development of large-wavelength
RT modes, and from that, understand some differences between the
structures of pairs of bubbles in PNs and in clusters.
\item Use the conditions for the development of the RT instability
to constrain the properties of the ambient density
to inflate large fat bubbles in PNs.
\item By using (1) and (2), try to account for the finding
that fat bubbles are rare in PNs, compared with their number
in clusters.
\item Discuss some effects due to instabilities:
($i$) The possibility of forming fast moving dense blobs.
($ii$) The destruction of the shell segments on the outermost
region (along the radial direction from the center of the system);
$(iii)$ Enable direct and efficient energy transfer from the
hot bubble interior to the ambient cooler gas, particularly
in clusters.
\item Guide future 3-dimensional numerical simulations in a search
for these instability modes.
\end{enumerate}

\section{THE ASSUMED FLOW STRUCTURE}
\subsection{Unperturbed Spherical Bubble}
I follow the derivation of Soker, Blanton, \& Sarazin (2002; hereafter
SBS).
The bubble is assumed to be inflated by a fast jet, which is shocked
and forms a spherical hot tenuous bubble, pushing a dense shell
into a constant-density medium.
Let $\dot E$ be the energy injection rate by the jet, $t$ the total
lifetime of the bubble from the start of the energy injection phase,
$\rho_c$ the undisturbed density of the ambient medium,
and $h$ the distance of the center of the bubble from the center of the
system (either the cluster or the PN).
The constant density assumption holds if the radius of the bubble
is small, $R_b \ll h$.
The justifications for these and others assumptions, e.g.,
neglecting the thermal pressure of the ambient medium
exterior to the bubble in clusters, are in SBS.
SBS, like Bicknell \& Begelman (1996), use the expression given by
Castor, McCray \& Weaver (1975) for the expansion of an interstellar
bubble (because the jet inflating the bubble has a finite width
this expression is not correct here when the bubble's radius is very small
and its density is high).
The radius of the bubble under these assumptions is given by
(Castor, McCray \& Weaver 1975)
\begin{equation}
R_b \simeq 0.76
\left(\frac {\dot E t^3}{\rho_c} \right)^{1/5}.
\end{equation}
The bubble expansion speed is
\begin{equation}
v_b=\frac{d R_b}{dt}= \frac {3}{5} \frac {R_b}{t},
\end{equation}
and its acceleration is
\begin{equation}
a_b=\frac{d v_b}{dt}= - \frac {6}{25} \frac {R_b}{t^2}.
\end{equation}
The accretion rate of ambient mass onto the expanding bubble's
shell is
\begin{equation}
\frac {d M_b}{dt} = 4 \pi R_b^2 v_b \rho_c ,
\end{equation}
and the total shell's mass is
\begin{equation}
M_b = \frac{4 \pi}{3} R_b^3 \rho_c .
\end{equation}

\subsection{Expansion into a Low Density Segment}
Instead of considering a realistic non-constant density
profile, which would require further assumptions to derive
analytical expressions, and then may obscure the
basic conditions I am aiming at, I follow the steps
of SBS in their section 6.
Instead of taking a monotonic decreasing density along the outward
radial direction, I assume that the bubble expands in a constant
density medium.
Then, when the spherical bubble reaches a radius $R_{b0}$ at
time $t_0$, a small segment on the side opposite to the center
of the system encounters a region of lower density $\rho_f$.
This segment, of solid angle $\Omega \ll 4 \pi$, expands faster
than the rest of the shell, into a ``funnel''.
Since the volume of the funnel is assumed to be small, the rest
of the bubble is assumed to expand like a spherical bubble, and
the pressure inside the bubble is like that of an unperturbed
spherical bubble.
The pressure $P$ inside the bubble acts on its surface,
such that the momentum changes according to
$d (M_b v_b)/dt=4 \pi R_b^2 P$.
At time $t_0$ the portion of the bubble's surface that enters
the low-density funnel has a mass $M_f = (\Omega/4 \pi) M_b$,
and velocity $v_f$.
The rate of change of the momentum of that mass is given by
$d (M_f v_f)/dt = \Omega R_b^2 P$.
 From the expression for the rate of change in momentum for the bubble
and the funnel-segment of mass I find
\begin{equation}
\frac {d(M_f v_f)}{dt} = \frac{\Omega}{4\pi} \frac{d( M_b v_b)}{dt}.
\end{equation}
Evaluating the two derivatives, using equations (2)-(5)
for the terms in the right-hand side, and
$d M_f/dt= \Omega R_f^2 v_f \rho_f$, where $R_f$ is the
distance of the funnel-segment from the center of the bubble,
gives for the segment's acceleration
\begin{equation}
a_f=\frac {dv_f}{dt} =
M_f^{-1} \left[ \frac{\Omega}{4 \pi} \frac {d(M_b v_b)}{dt} -
v_f \frac {dM_f}{dt} \right] \\ \nonumber
=\frac{\Omega}{M_f} \left[\frac{7}{9}-
\left( \frac{R_f v_f}{R_b v_b} \right)^2 \frac{\rho_f}{\rho_c}
\right]  v_b^2 R_b^2 \rho_c .
\end{equation}
Close to the time the bubble encounters the low density funnel,
i.e., $t-t_0 \ll t_0$, we can take the values at $t=t_0$:
$M_f=\Omega M_b/4\pi$, and $v_f=v_b$.
Substituting these in the last equations, and using
equation (5) for $M_b$ and equation (2) for $v_b$,
gives
\begin{equation}
a_{f0}=\frac{3}{25} \frac{R_b}{t^2} \left(7-9 \frac{\rho_f}{\rho_c}
\right).
\end{equation}

\section{RAYLEIGH-TAYLOR INSTABILITY}
\subsection{Planetary Nebulae}
In PNs the thermal pressure of the ambient medium can be neglected
(before photoionization by the central star starts).
The low density bubble interior exerts force on the dense shell.
The interface will be RT unstable when the acceleration as
written in equations (7) or (8) is positive $a_{f0}>0$,
namely the acceleration is in the opposite direction to
the density gradient. This gives the condition for instability
\begin{equation}
\frac {\rho_f}{\rho_c} < \frac{7}{9}
\end{equation}
I examine the outermost region (relative to the center of the
nebula) of the bubble, which is the least stable region
because of the steeper density gradient.
For a medium formed by an asymptotic giant branch (AGB) star
losing mass at a constant rate and velocity, the density profile
along the radial direction from the nebular center along the bubble's
center is
$\rho \propto (h+R_b)^{-2}$, where $h$ is the distance of the bubble's
center from the central star.
This gives at time $t_0$ and with the assumptions used
the previous section, i.e., the definition of $\rho_f$,
\begin{equation}
\frac {\rho_f}{\rho_c} = \left(\frac {h}{h+R_b} \right)^2.
\end{equation}
The condition for RT instabilities to develop becomes
\begin{equation}
\frac {R_B}{h} > \left(\frac{9}{7}\right)^{1/2}-1 = 0.13.
\end{equation}

The growth time for RT instabilities (the e-folding  time) is given
by $\tau_{\rm RT} = | a_f k |^{-1/2}$,
where $\lambda$ and $k=2 \pi /\lambda$ are the wavelength and
wavenumber of the mode being considered.
Substituting the acceleration $a_{f0}$ for the RT unstable case gives
\begin{equation}
\tau_{\rm RT} = \left( \frac{25}{6 \pi} \right)^{1/2}
\left( \frac{\lambda}{R_b} \right)^{1/2}
\left(7-9 \frac{\rho_f}{\rho_c} \right)^{-1/2} t.
\end{equation}
For a wavelength that completely disrupts the bubble
$\lambda=2 R_b$, and with the density profile given
by equation (10), the growth time becomes shorter than
the bubble age, i.e., $\tau_{\rm RT} <t$, when the bubble radius
becomes $R_B\simeq h/2$.
Shorter wavelengths, which will also destruct the front of the
bubble, $0.5 R_b \lesssim \lambda \lesssim 2 R_B$, will grow
faster.

Despite the simplifying assumptions made in deriving the expression
for the acceleration of the funnel segment (eqs.7-8), and the
growth time of the unstable mode given by equation (12),
the condition above seems robust:
If the ambient medium in a PN (or a proto-PN) was formed
by a progenitor AGB star whose mass loss rate and speed was
constant, then bubbles, if formed, become unstable before they
grow beyond radii of $R_b \sim 0.2-0.5 h$.
No large and well defined close bubbles are expected in this case.
Also, the outer front will expand much faster, and the bubble will
lose its more or less spherical structure.
With a constant mass loss rate from the AGB progenitor, jets can
inflate bubbles (for the conditions on the jets see paper-II),
but either they will be small bubbles far from the center
or they will have a destructed outer front.
To form large and well defined closed bubbles, as in the Owl
Nebula (NGC 3587, Guerrero et al.\ 2003) or VV 171 (Sahai 2000),
the mass loss history from the AGB progenitor must be different.
Namely, the density should not decline too fast with radius.
This implies that the mass loss rate was very high, and then
declined slowly, such that the bubble was inflated in a region
where the density profile was very shallow as in clusters
(see next subsection).
Eventually instability will develop; in the Owl Nebula
the shell is already fragmented to some degree.

\subsection{Clusters}

SBS study the RT instability of the interface between the
bubble interior and the dense shell in clusters, where
gravitational acceleration is important.
For bubbles inflated about a center which is not the cluster center,
the least stable portion of the bubble is its outermost
region because the dense shell lies above the low-density
interior of the bubble.
If, for example, the acceleration due to the interior pressure
vanishes, this region in clusters is unstable because of
the gravitational field (SBS).
The outermost region is also the one most prone to
RT instability because of the steeper density profile,
as shown in the previous subsection.
The condition for RT instability to develop at the outermost
region of bubbles in clusters because of gravity alone is given
by equation (25) in SBS, which is for the density and gravity
of Abell~2052, taken from Blanton et al.\ (2001).
For an energy injection rate of $\dot E= 3 \times 10^{44} \erg \s^{-1}$
the instability condition, which is that the gravitational
acceleration exceed that of the bubble deceleration, reads
\begin{equation}
R_b + h \la 15
\left(\frac {R_b}{10 \, {\rm kpc}} \right)^{1.4}
\, {\rm kpc} \, ,
\end{equation}
where, as before, $h$ is the distance of the center of the
bubble from the center of the cluster.
Basically, the outermost region of bubbles in clusters
becomes unstable because of gravity alone when the bubble
is relatively large.
For the above parameters, for example, as long as
$h <20 \kpc$ the bubble becomes unstable only
when $R_b > h$.
SBS find also that after the bubble becomes unstable,
the growth time for RT instabilities is comparable
to the estimated age of the bubbles.
When only gravity is considered, and not density gradients,
this implies that in most cases the bubbles will be stable during
their inflation phase, while in other cases the bubbles will start to
be fragmented, but only in the outer regions.
SBS argue that this is in reasonable agreement with the fact that
the bubbles in Abell~2052 are fairly complete, except for possible
gaps at their outer edges  (the north side for the north bubble
and the south side for the south bubble).

Because gravity influences instability in clusters only when
the bubble is large, I consider the effect of the density gradient
on instability in the outermost region of the bubble while
neglecting the effect of gravity; condition (9) is applicable.
Because the density profile in clusters is much shallower than
the one given by equation (10), the bubble must become much
larger to become unstable in its outermost region, compared
with the condition given in equation (11).
By examining the density profiles of two clusters,
Abell~2052 (fig. 2 of Blanton et al.\ 2001),
and Hydra~A (fig. 3 of David et al.\ 2001), I find
that condition (9) for instability at the outermost segment
of the shell is fulfilled under the following conditions.
In Hydra~A the electron density profile in the range
$10 \lesssim r \lesssim 30 \kpc$ can be approximated by
$n_e = 0.06 (r/10 \kpc)^{-0.6} \cm^{-3}$, being much shallower
for $r < 10 \kpc$, where $r$ is the distance from the
cluster center.
This implies, from equation (9), that for
$10 \lesssim r \lesssim 20 \kpc$ the outermost front of the bubble
becomes unstable when $R_b \gtrsim 0.5 h$.
If the center of the bubble is further in,
$h < 10 \kpc$, the ratio $R_b/h$ has to become larger
for instability to start.
In Abell~2052 the density profile is influenced a lot by
the shells around the two bubbles, so it is hard to tell
what was the density before the bubbles were inflated.
In any case, in Abell~2052 the density is flat up to
$R \sim 20 \kpc$, and then declines as $r^{-1}$.
The average slope between $ r\simeq 10 \kpc$ and $r \simeq 40 \kpc$
is $r^{-0.6}$, similar to that in Hydra~A.
Since in Abell~2052 the bubbles centers are at $h < 20 \kpc$,
the bubbles radii must becomes comparable to $h$ for
the instability to develop.

The conclusion from the above discussion is that the outermost
segments of bubbles in clusters of galaxies become unstable,
for bubbles inflated close to the center, only when
\begin{equation}
\frac {R_B}{h} \gtrsim 1 .
\end{equation}
This is similar to the condition derived when gravitational
acceleration is considered.

\section{DISCUSSION}

The analytical analysis of the previous sections,
when combined with results from earlier papers
where similar analytical analyses were used
(paper-I, paper-II, and SBS), leads to some useful conclusions.
As the main aim is to compare the properties of pairs of
bubbles in clusters with those in PNs, the approximations used
and assumptions made affect in similar ways the results
for both classes of objects.
Therefore, despite these simplifications the following conclusions
regarding the comparison are robust, with the advantage that
the analytical approach clarifies the roles of the different
processes, e.g., effect of gravity and steep density profiles
in clusters.
\begin{enumerate}
\item The comparison of the instability criterion in clusters,
equation (14), with that for PNs, equation (11), teaches us
the following.
Bubbles in most PNs are destructed when they are small.
Only when the mass loss rate from the progenitor AGB stars
decreases for a long enough time period, but still is high and slow,
does the ambient medium favor the formation of a stable large bubble.
Adding to that the constraint on the jet's properties derived
in paper-II, that the jet be wide-angle or precess, we can
understand why fat bubbles in PNs are rare.
In clusters, on the other hand, the bubbles become unstable only
when they are large, and then they can survive for
a relatively long time (SBS).
{{{ Strong magnetic fields inside bubbles in clusters will stabelize
them further (Robinson et al.\ 2004). }}}
\item The outermost region of the bubbles (the center of the bubble
is displaced by $h$ from the center of the system)
is the most unstable region, because the density gradient is the larger
there. The shell segment in that region will be the first to become
unstable, and destructed. Eventually it may disappear.
However, because this region is along the line of the center of the
bubble and the center of the system, any expanding narrow jet that
may be blown after the bubble was inflated may also remove that region.
A clumpy region and no indication of a jet will hint at instabilities,
as in the bubbles of Abell~2052.
\item From equation (8) it turns out that the outermost region of
the shell will be accelerated to higher velocities, and the
structure will change from a spherical bubble to an elongated one
(a lobe). This effect will be more significant in PNs.
Elongated lobes in PNs will also be formed by narrow jets
(Soker 2002;  Lee \& Sahai 2003; paper-II).
I expect that in most PNs that have elongated lobes, whether closed
or open at their outermost region, the lobes are formed by fast narrow
jets, rather than from spherical bubbles which expanded along the
radial direction.
Again, nice spherical (fat) bubbles are rare in PNs.
\item The short-wavelength RT instability modes grow on
short time scales.
As the bubble's outermost region becomes unstable,
these dense and small blobs are likely to be formed.
These dense blobs may continue to move fast even after the
the rest of the shell is decelerated as it expands and its density
decreases. Such blobs can plow through the shell, and appear as
small blobs$-$`bullets'$-$ahead of the main shell
in the outer, i.e., radial, direction.
This process is more likely to occur in PNs, where the bubbles
become unstable when they are small and expand fast.
Theses dense blobs may be the ballistic `bullets' which were suggested
as the source of the strings in the bipolar nebula,
i.e., made of two lobes, of the massive star $\eta$ Carinae
(Redman, Meaburn, \& Holloway 2002).
The strings in $\eta$ Carinae are  radial, straight, narrow and long
optical filaments emerging from the inner nebula$-$the Homunculus
(Weis, Duschl, \& Chu 1999).
Redman et al.\ (2002) suggest that the strings are the decelerating
flow of ablated gas from these bullets.
Here I propose the RT instability during the very early
inflation phase of the two lobes of the Homunculus as the
source of these bullets.
\item
Heat may be conducted from the hot bubble interior to the
cooler surrounding medium.
In the past this process was studied in more detail for PNs
(Soker 1994; Zhekov \& Myasnikov 2000; Soker \& Kastner 2003).
This process, among other things, was claimed to be able to heat
somewhat the cooler ($\sim 10^ 4\K$) environment in PNs.
The heating will not be isotropic because it is regulated by
magnetic fields (Soker 1994; Zhekov \& Myasnikov 2000).
The magnetic field lines should connect the hot interior with the
cooler surrounding, or else they inhibit efficient heat conduction.
The RT instability may mix the two media, causing the magnetic
fields in the two media to reconnect, thereby allowing
efficient heat conduction.
This in itself may contribute to the destruction of the dense and
somewhat cooler shell around the bubble.
\end{enumerate}

\acknowledgements
This research was supported in part by the Israel Science Foundation.

\end{document}